\documentclass[aps,prl,reprint,showpacs,amsmath,amssymb]{revtex4-1}
\usepackage{graphicx}
\usepackage{ifpdf}

\usepackage[dvips,usenames]{color}
\begin{document}

\title{Duality in multi-channel Luttinger Liquid with local scatterer}


\author{Igor V.\ Yurkevich}
\affiliation{Nonlinearity and Complexity Research Group, Aston University, Birmingham B4 7ET, United Kingdom}

\begin{abstract}
We have devised a general scheme that provides exact results for the scaling dimensions of all allowed in multi-channel Luttinger Liquids local perturbations. The technique developed provides universal description and establishes relations between scaling dimensions of perturbations in different phases of Luttinger Liquid. These multiple relations between scaling dimensions are universal. They are valid for all Luttinger Liquids with arbitrary current-current and density-density interactions. We have also established duality transformation between different subsets of Luttinger Liquid phases known for single-channel liquid as duality between weak link and weak scatterer limits.
\end{abstract}

\maketitle
Solid state systems like Fractional Quantum Hall systems\cite{FQH}, Quantum Spin Quantum Hall or topological insulators \cite{TeoKane, TI}, carbon nanotubes \cite{CN} and quantum wires are now routinely described in terms of few (chiral or not) channels with intra- and inter-channel interactions. Similar description is also applied to cold atoms mixtures \cite{mixture}, ballistic quasi-one-dimensional waveguides \cite{Y4}, hollow-core fibers \cite{Fibers} and one-dimensional electron-phonon systems \cite{Y, Yu}. The intra-channel interactions being taking into account lead to formation of Luttinger Liquid \cite{LL} for each individual channel. The perturbations that scatter or tunnel particles within the same channel are either irrelevant (scaling dimension is higher than physical dimension which is unity for a local perturbation) or relevant (the scaling dimension is lower than one) in terms of renormalization group analysis \cite{KF}. It means that a single channel is to be found in one of two states (depending on material parameters): perfectly conducting or insulating.\\

The inter-channel interactions make scaling dimensions of all perturbation inter-dependent. A new and rich phase diagram emerges as the result. Since interaction in a single channel makes it either perfectly conducting or completely insulating the best starting point to examine effect of local scatterer is to assume that the state of $N$-channel liquid has $n$ insulating and ($N-n$) conducting channels. Each particular realization of such configuration is called a phase. This is natural generalization of phases observed in two-channel problems (like spin and charge channels). The phase is stable if all allowed perturbations are irrelevant. The condition that all scaling dimensions of local perturbations are higher than one (irrelevant perturbations) defines a region of physical parameters where this particular phase can be observed. Intersections of different regions correspond to unstable fixed points meaning that there is no unique phase for those system parameters and which phase is realized depends on bare values of perturbations (multiple attraction basins). On the other hand if the union of all those regions does not cover the whole space it means that there is a range of system parameters where none of bare phases is stable, there must be a new stable fixed point corresponding to a new phase of matter. Such situation is known to occur for one dimensional electrons with spin but without $SU(2)$ symmetry \cite{KF} or in topological insulators at strong interactions \cite{TeoKane}. It occurs that when two-particle local scattering is taken into account all bare phases become unstable in some region of material parameters.\\

The scaling dimensions of two operators defining instability of two phases of a single-channel Luttinger Liquid are known to be inversely proportional to each other \cite{KF}. This relation between scaling dimensions is consequence of the duality between weak and strong scatterer limits. Recently it was also observed in \cite {Y} that coupling of interacting electrons (the Luttinger Liquid) to acoustic phonons did not change the duality. The fact that scaling dimensions of the operators determining instability of fermionic channel were changed essentially but stayed inversely proportional to each other came as surprise. The scaling dimensions were derived explicitly as the result of lengthy calculations and the reason behind duality was unclear. It will be shown in this Letter that the duality observed in \cite{Y} is just one particular example of universal property which is a generalized duality relation valid for arbitrary multi-channel Luttinger Liquid.\\

We develop a generic scheme of calculation of scaling dimensions of all symmetry allowed local perturbations not just because it gives a machinery of dealing with multi-component phases. We do it to reveal hidden symmetries reflected in relations between scaling dimensions of perturbations in different phases. We will show below that the matrix ${\hat\Delta}$ which has scaling dimensions of different perturbations as entries is given by
\begin{align}
{\hat\Delta}={\hat\Delta}_{I}\oplus{\hat\Delta}_C
\end{align}
for each phase characterized by a set of insulating and the complimentary set of conducting states. The introduction of two sectors for multi-channel Luttinger Liquid is natural generalization of accepted now in two-channel problems nomenclature: $II$ or $CC$ stand for both channels being insulating or conducting, $CI$ and $IC$ for one conducting and one insulating channel. The matrices ${\hat\Delta}_{{\rm I}}$ and ${\hat\Delta}_{{\rm C}}$ are defined on two orthogonal subspaces of insulating ($I$) and conducting ($C$) channels. Their matrix elements are not independent though. To construct them one has to find matrix ${\hat\Delta}_{\theta}$ of scaling dimensions in ideally conducting phase. The projection of this matrix onto $I$-subspace and inverting the projection gives ${\hat\Delta}_{{\rm I}}$. The projection of the inverse matrix ${\hat\Delta}_{\theta}^{-1}$ and following inversion of the projection gives ${\hat\Delta}_{\rm C}$. The common source for both matrices ${\hat\Delta}_{\rm I}$ and ${\hat\Delta}_{\rm C}$ implies relations between scaling dimensions in conducting and insulating sectors. These relations impose restrictions on inter-phase boundaries in material parameter space and can be used to predict (without system dependent calculations) whether a new phase is expected or not on quite generic basis.\\

Multi-channel Luttinger Liquid is a set of $N$ interacting individual liquids. Each channel is labeled with an index $i=1, ..., N$ and described in terms of density, $n_i=\partial_x\theta_i/\pi$, and current, $j_i=\partial_x\phi_i/\pi$, fields. Each channel is characterized by its own velocity $v_i$ and the Luttinger parameter $K_i$ reflecting strength and statistics of underlying particles. The Lagrangian written in terms of bosonic fields $\theta_i(x,t)$ and $\phi_i(x,t)$,
\begin{align}\label{L}
{\cal L}=\frac{1}{\pi}\theta_i\partial_t\partial_x\phi_i-H[\theta, \phi]\,,
\end{align}
contains Hamiltonian part which is (neglecting backscattering) is a sum of two quadratic forms for density-density and current-current interactions:
\begin{align}\label{H}
H[\theta,\phi]=\frac{1}{2\pi}\partial_x\theta_i\left(\frac{v_i}{K_i}\,\delta_{ij}+g^{\theta}_{ij}\right)\partial_x\theta_j\nonumber\\
+\frac{1}{2\pi}\partial_x\phi_i\left(v_i\,K_i\,\delta_{ij}+g^{\phi}_{ij}\right)\partial_x\phi_j\,.
\end{align}
The diagonal terms describe individual channels while inter-channel interactions are included into interaction matrices ${\hat g}^{\theta}$ and ${\hat g}^{\phi}$. There are two complementary representations either in terms of density, $\theta^{\rm T}=(\theta_1, ..., \theta_N)$, and current, $\phi^{\rm T}=(\phi_1, ..., \phi_N)$, vector fields or in terms of chiral right-moving, $\theta_{R}=\phi+\theta$, and left-moving, $\theta_{L}=\phi-\theta$, fields. We will be switching between these two representations because they both have advantages when performing different tasks. The first step will be the reduction of the Lagrangian to a diagonal form and it is much easier in $(\theta\,,\phi)$ representation because we have the Hamiltonian part which is sum of two quadratic forms in this representation. The transformation matrices
\begin{align}\label{M}
\theta= {\hat M}_{\theta}{\tilde\theta},\quad \phi= {\hat M}_{\phi}\,{\tilde\phi}
\end{align}
must diagonalize Hamiltonian part of the action and also preserve the structure of the first term in Eq. (\ref{L}) (which is the same as the preservation of the commutation relations in the operator formulation). The later requirement is imposing connection between those transformations:
\begin{align}\label{restriction}
{\hat M}_{\theta}^T\, {\hat M}_{\phi}=1\,.
\end{align}
These transformations always exist and can be constructed as four-steps procedure with each step preserving the scalar product $\theta^{\rm T}\phi={\tilde\theta}^{\rm T}\,{\tilde\phi}$. First, one can apply unitary transformation to diagonalize the quadratic form in $\phi$ and simultaneously apply the same unitary transformation to $\theta$-vector. Second, one rescales each component of the new $\phi$-field to absorb the eigenvalues and turn the quadratic form into a scalar product while $\theta$-vector is subject to inverse rescaling. Now we may again apply identical unitary transformations to both vectors choosing it to diagonalize the quadratic form in the new $\theta$-fields (the kernel stays real and symmetric during all transformations). Finally, we can rescale each component of $\theta$- and inversely rescale $\phi$-vectors in such way that the coefficients in front of either $i$-th component are the same, $u_i$ (they will be new velocities). The resulting Lagrangian in terms of new fields is given by the expression:
\begin{align}
{\cal L}=\frac{1}{\pi}{\tilde\theta}_i\partial_t\partial_x{\tilde\phi}_i-
\frac{u_i}{2\pi}\,\left[\left(\partial_x{\tilde\theta}_i\right)^2+\left(\partial_x{\tilde\phi}_i\right)^2\right]\,.
\end{align}
In a translational invariant system the transformations (\ref{M}) relate the Green functions of interacting Luttinger Liquids and the Green functions of uncoupled liquids. Since the latter is well known one can easily find, for example, local Green functions $iG^{ij}_{\theta}(t;t')=\langle\theta_i(x,t)\,\theta_j(x,t')\rangle$ and $iG^{ij}_{\phi}(t;t')=\langle\phi_i(x,t)\,\phi_j(x,t')\rangle$. In matrix form the retarded components can be written as
\begin{align}
{\hat G}^R_{\theta}(\omega)=-\frac{i\pi}{2}\,\frac{{\hat\Delta}_{\theta}}{\omega+i0}\,,\quad
{\hat G}^R_{\phi}(\omega)=-\frac{i\pi}{2}\,\frac{{\hat\Delta}_{\phi}}{\omega+i0}\,,
\end{align}
with matrices
\begin{align}
{\hat\Delta}_{\theta}={\hat M}_{\theta}\,{\hat M}^{\rm T}_{\theta}\,,\quad
{\hat\Delta}_{\phi}={\hat M}_{\phi}\,{\hat M}^{\rm T}_{\phi}
\end{align}
being inversely proportional to each other
\begin{align}
{\hat\Delta}_{\theta}\,{\hat\Delta}_{\phi}=1\,.
\end{align}

In the presence of a scatterer at the origin $x=0$ the transformation (\ref{M}) should be performed on the left and on the right of the scatterer separately because the fields are no longer continuous across the origin. To take into account boundary conditions that relate those fields it is now convenient to switch to right- and left-movers combining them into incoming ($in$) and outgoing ($out$) fields:
\begin{align}
\Theta_{out}=\left(\begin{array}{c}\theta_R(+0,t) \\\theta_L(-0,t) \\\end{array}\right)\,,\quad
\Theta_{in}=\left(\begin{array}{c}\theta_L(+0,t) \\\theta_R(-0,t) \\\end{array}\right)\,.
\end{align}
The same definition is used to construct new transformed fields ${\tilde\Theta}_{in}$ and ${\tilde\Theta}_{out}$. The boundary conditions for the original and new fields can be written using $S$-matrix
\begin{align}
\Theta_{out}={\hat S}\,\Theta_{in}\,,\quad {\tilde\Theta}_{out}={\hat {\tilde S}}\,{\tilde\Theta}_{in}\,.
\end{align}
While the scattering matrix for the new fields has yet to be found the scattering matrix for the original fields is known and it depends on the bare phase of Luttinger Liquid whose stability against scattering we would like to try. The phase under investigation can be presented by a set of reflection coefficients $R_i\,,\,\,i=1,...,N$ where $R_i=0$ if the $i$-th channel is ideally transparent (conducting) or $R_i=1$ if it is fully blocked (insulating). Then the scattering matrix can be written as
\begin{align}
{\hat S}=\left(
           \begin{array}{cc}
             {\hat R} & {\hat T} \\
             {\hat T} & {\hat R} \\
           \end{array}
         \right),\,\, {\hat R}={\rm diag}(R_1,...,R_N)\,.
\end{align}
The current conservation $\sum_i\Theta^i_{in}=\sum_i\Theta^i_{out}$ requires that ${\hat R}+{\hat T}=1$. It is important to stress that the scattering matrix we use is the one for current/density fields and cannot be used to describe an arbitrary scattering of real particles whose excitations are described in terms of the Luttinger Liquids. The matrices ${\hat R}$ and ${\hat T}$ are nothing but projectors on the subspaces of insulating ($I$) and conducting ($C$) channels. \\

Our task now is to derive correspondence between original and new scattering states given the transformation Eq.(\ref{M}). The derivation is straightforward and leads to the following relations
\begin{align}\label{rel}
\Theta_{out}=\left[{\hat{\cal M}}_+{\hat{\tilde S}}+{\hat{\cal M}}_-\right]{\tilde\Theta}_{in}\,,\\\nonumber
\Theta_{in}=\left[{\hat{\cal M}}_-{\hat{\tilde S}}+{\hat{\cal M}}_+\right]{\tilde\Theta}_{in}\,,
\end{align}
whose consistency defines the new scattering matrix in terms of the original one:
\begin{align}\label{S}
{\hat{\tilde S}}=\left[{\hat{\cal M}}_+-{\hat S}{\hat{\cal M}}_-\right]^{-1}{\hat S}\,\left[{\hat{\cal M}}_+-{\hat S}{\hat{\cal M}}_-\right]\,.
\end{align}
Here the extended ($2N\times 2N$) block diagonal matrices contain identical blocks ${\hat{\cal M}}_{\pm}={\rm diag}({\hat M}_{\pm},{\hat M}_{\pm})$ which are linear combinations of the transformation matrices, ${\hat M}_{\pm}=\left[{\hat M}_{\phi}\pm{\hat M}_{\theta}\right]/2$.\\

The expression (\ref{S}) for the scattering matrix can be simplified noticing that the transformation in the right/left-movers space
\begin{align}
{\hat\Theta}_{in}\to {\hat L}{\hat\Theta}_{in}\,,\quad
{\hat L}=\frac{1}{{\sqrt 2}}\left(
  \begin{array}{cc}
    1 & 1 \\
    -1 & 1 \\
  \end{array}
\right)
\end{align}
makes ${\hat S}$ and, therefore, ${\hat{\tilde S}}$ block diagonal
\begin{align}
{\hat L}^{-1}{\hat S}{\hat L}=\left(
                           \begin{array}{cc}
                             {\hat R}-{\hat T} & 0 \\
                             0 & 1 \\
                           \end{array}
                         \right)\,,\quad
{\hat L}^{-1}{\hat{\tilde S}}{\hat L}=\left(
                           \begin{array}{cc}
                             {\hat s} & 0 \\
                             0 & 1 \\
                           \end{array}
                         \right)\,,
\end{align}
with the only nontrivial block entry
\begin{align}
{\hat s}={\hat K}^{-1}({\hat R}-{\hat T})\,{\hat K}\,,\quad {\hat K}={\hat R}{\hat M}_{\theta}+{\hat T}{\hat M}_{\phi}\,.
\end{align}
The rotation with the matrix ${\hat L}$ is equivalent to the redefinition of right- and left-moving incoming fields in terms of new uncorrelated fields $\theta^{\pm}_{in}=\theta_{R\,in}\pm\theta_{L\,in}$. \\

To find scaling dimensions of perturbations it is sufficient to know the correlation functions of incoming fields (the rest can be restored using the scattering matrix, if necessary). The Green function of new incoming fields is essentially trivial because the scatterer is located 'downstream' for incoming fields and there is no interaction between incoming and outgoing fields. The Green function $i{\hat{\tilde G}}_{in}=\langle{\tilde\Theta}_{in}\,{\tilde\Theta}^{\rm T}_{in}\rangle$ is simply given by the expression valid in translational invariant problem
\begin{align}
{\hat{\tilde G}}_{in}^R(\omega)=-\frac{2\pi\,i}{\omega+i0}\,{\hat 1}\,.
\end{align}
The Green functions of incoming original, ${\hat G}_{in}$, and new, ${\hat{\tilde G}}_{in}$, fields are related to each other (Eq. (\ref{rel})):
\begin{align}\label{G}
{\hat L}^{-1}{\hat G}_{in}^R(\omega){\hat L}=-\frac{2\pi i}{\omega+i0}\,\left(
                                                                         \begin{array}{cc}
                                                                           {\hat\Delta} & 0 \\
                                                                           0 & {\hat\Delta}_{\phi} \\
                                                                         \end{array}
                                                                       \right)\,.
\end{align}
Here the matrix ${\hat\Delta}$ is a direct sum of two matrices, each being projected onto $I$ or $C$ subspaces:
\begin{align}
{\hat\Delta}={\hat R}\,{\hat\delta}^{-1}\,{\hat R} + {\hat T}\,{\hat\delta}^{-1}\,{\hat T}\,.
\end{align}
The matrix ${\hat\delta}$ has similar structure of a direct sum of projected matrices:
\begin{align}
{\hat\delta}={\hat R}\,{\hat\Delta}_{\theta}\,{\hat R} + {\hat T}\,{\hat\Delta}_{\phi}\,{\hat T}.
\end{align}
The inverse matrix is also direct sum with elements defined in two orthogonal to each other subspaces $I$ of insulating channels (with the projector ${\hat R}$) and $C$ of conducting channels (with the projector ${\hat T}$):
\begin{align}
{\hat\Delta}={\hat\Delta}_{\rm I}\oplus{\hat\Delta}_{\rm C}
\end{align}
These matrices are defined as follows. To construct ${\hat\Delta}_{\rm I}$ one has to project ${\hat\Delta}_{\theta}$ onto subspace $I$ of insulating channels or, in other words, consider all non-zero matrix elements of the matrix ${\hat R}\,{\hat\Delta}_{\theta}\,{\hat R}$ because ${\hat R}$ is the projector onto that subspace. The inversion of the projected matrix also belongs to the $I$-subspace of insulating channels and is called ${\hat\Delta}_{\rm I}$. Analogously, to construct ${\hat\Delta}_{\rm C}$ one has to project ${\hat\Delta}_{\phi}$ onto $C$-subspace of conducting channels or, in other words, consider all non-zero matrix elements of the matrix ${\hat T}\,{\hat\Delta}_{\phi}\,{\hat T}$ because ${\hat T}$ is the projector onto that subspace. The inversion of the projected matrix also belongs to the subspace $C$ of conducting channels and is called ${\hat\Delta}_{\rm C}$.\\

There is obvious duality: interchanging ${\hat R}\leftrightarrow{\hat T}$ and $\theta\leftrightarrow\phi$ equivalent to interchange
\begin{align}
{\hat\Delta}_{\rm C}\leftrightarrow{\hat\Delta}_{\rm I}\,.
\end{align}
This is the generalization of the well-known in single-channel problem duality between weak link (insulating phase) and weak scatterer (conducting phase). In a generic situation of multi-channel Luttinger Liquid the generalized duality relates two phases: one of them has $n$ insulating and $N-n$ conducting channels, another ('partner') phase has $N-n$ insulating and $n$ conducting channels. The simplest example is perfectly conducting phase (all channels are conductors) with ${\hat\Delta}_{\rm cond}={\hat\Delta}_{\theta}$ against perfectly insulating phase (all channels are insulators) with ${\hat\Delta}_{\rm ins}={\hat\Delta}_{\phi}={\hat\Delta}_{\rm cond}^{-1}$.\\

This duality connects some 'partner' phases but does not relate scaling dimensions of similar perturbations in seemingly unrelated phases. Nevertheless, as we will show below there are intimate relations between scaling dimensions beyond the one discussed above. We will leave general analysis for future investigations. In this Letter we will focus on two-channel liquid to reveal universality of the relation between scaling dimensions which was observed in \cite{Y}. \\

To analyze stability of a phase, first of all, we have to parameterize perturbations in terms of the incoming fields. It is easy since we provide description of stability of the phases where each channel is either perfectly conducting or insulating and all inter- and intra-channel particle transfers can be written in terms of incoming fields only. An arbitrary process of simultaneous transfer of few particles with $n_{ij}$ being the number of particles transferred from channel $i$ to channel $j$ is described by the perturbation
\begin{align}
T=v\,\cos\left[\sum_{ij}n_{ij}\left(\theta^{-}_{ij}+\theta^{+}_{ij}\right)\right]\,,
\end{align}
where $\theta^{+}_{ij}=\theta_i^+-\theta_j^+$ and $\theta^{-}_{ij}=\pm\theta_i^-\pm(R_j-T_j)\theta_j^-$.
The scaling dimension of this perturbation is given by the sum of two dimensions since $\theta^{-}_{ij}$ and $\theta^+_{ij}$ are independent of each other. According to Eq. (\ref{G}) the scaling dimension related to $\theta^+_{in}$ is the same for all phases while the other one is governed by the matrix ${\hat\Delta}$ and depends on the phase which is uniquely defined by the reflection matrix ${\hat R}$.\\

For a particular problem of two-channel Luttinger Liquid with no particle transfers between channels due to distinct nature of the particles (fermions and phonons in the paper \cite{Y}) the only perturbations allowed are intra-channel ones. The most generic perturbation intra-channel transfer $T\sim \cos\left[n_1\theta_{in}^1+n_2\theta_{in}^2\right]$ has the scaling dimension
\begin{align}
D_{n_1\,n_2}(R_1, R_2)=n_1^2\,\Delta_{11}+n_2^2\,\Delta_{22}+2n_1 n_2\,\Delta_{12}
\end{align}
that depends on the phase which is tried for stability. The phase is uniquely described by the diagonal matrix elements of the matrix ${\hat R}={\rm diag}(R_1, R_2)$. The scaling dimension of the phase where both channels are conducting ($CC$-phase corresponding to $R_1=R_2=0$) is given by
\begin{align}\label{CC}
D_{n_1\,n_2}(0, 0)=n_1^2\,\Delta_{\theta}^{11}+n_2^2\,\Delta_{\theta}^{22}+2n_1 n_2\,\Delta_{\theta}^{12}\,.
\end{align}
The $II$-phase ($R_1=R_2=1$) is restored by the duality
\begin{align}\label{II}
D_{n_1\,n_2}(1, 1)=n_1^2\,\Delta_{\phi}^{11}+n_2^2\,\Delta_{\phi}^{22}+2n_1 n_2\,\Delta_{\phi}^{12}\,.
\end{align}
For $IC$- ($R_1=1, R_2=0$) and by duality for $CI$-phases ($R_1=0, R_2=1$) we get
\begin{align}\label{IC}
D_{n_1\,n_2}(1, 0)=\frac{n_1^2}{\Delta_{\theta}^{11}}+\frac{n_2^2}{\Delta_{\phi}^{22}}\,,\,\,\,\,
D_{n_1\,n_2}(0, 1)=\frac{n_1^2}{\Delta_{\phi}^{11}}+\frac{n_2^2}{\Delta_{\theta}^{22}}\,.
\end{align}

In the paper \cite{Y} only one-particle perturbations in fermion channel ($n_1=1,\,n_2=0$) were considered. The second channel did not contained renormalizable perturbation since phonon scattering is given by a local quadratic perturbation which at low energy leads to either $R_2=0$ (local mass distortion) or $R_2=1$ (local pinning).  The weak scatterer and weak link scaling dimensions were defined as
\begin{align}
\Delta_{ws}(R_2)=D_{10}(0,R_2)\,,\quad \Delta_{wl}(R_2)=D_{10}(1,R_2)\,.
\end{align}
One can see from Eqs. (\ref{CC}-\ref{IC}) that
\begin{align}
\Delta_{ws}(R_2)\,\Delta_{wl}(R_2)=1
\end{align}
and the duality observed in \cite{Y} is just one particular case of general relations Eqs. (\ref{CC}-\ref{IC}). \\

In conclusion, we have devised new approach to calculation of scaling dimensions of the most generic local perturbations in multi-channel Luttinger Liquids. We have derived explicit expression for the matrix of scaling dimensions for all phases parameterized by the number of conducting and insulating channels. We have demonstrated symmetry with respect to simultaneous interchange of the current and density fields and subspaces of insulating and conducting channels (known as duality in one-channel case). We have also explained on general grounds (without even specifying inter- and intra-channel interactions) the duality between scaling dimensions of one-particle perturbations in two-channel liquids making obvious the result observed in \cite{Y}. The knowledge of perturbations scaling dimensions allows to construct complete phase diagram. Although analysis of numerous practically important realizations was beyond the scope of this Letter we believe that the scheme will prove very useful and will be used to determine existence of new stable phases of multi-channel Luttinger Liquids.

\begin{acknowledgments}
I would like to thank I. V. Lerner and O. M. Yevtushenko for stimulating and fruitful discussions and V. E. Kravtsov for critical reading of the manuscript.
\end{acknowledgments}


\end{document}